\documentclass[aps,pra,twocolumn,showpacs,floatfix]{revtex4}

\usepackage{amsmath,amssymb,float,graphicx}

\begin{document}

\title{Comment  on ``Sound velocity and multibranch Bogoliubov spectrum
of an elongated Fermi superfluid in the BEC-BCS crossover"}

\author{P. Capuzzi}
\email{capuzzi@sns.it}
\author{P. Vignolo}
\email{vignolo@sns.it}
\author{F. Federici}
\email{fr.federici@sns.it}
\author{M. P. Tosi}
\email{tosim@sns.it}
\affiliation{NEST-CNR-INFM, Scuola Normale Superiore, I-56126 Pisa, Italy}

\begin{abstract}
The work  by  T. K. Ghosh and K. Machida [\texttt{cond-mat/0510160}
and Phys. Rev. A \textbf{73}, 013613 (2006)] on the sound velocity
in a cylindrically confined Fermi superfluid obeying a power-law
equation of state is shown to make use of an improper projection of
the sound wave equation. This inaccuracy fully accounts for the
difference between their results and those previously reported by Capuzzi
\textit{et al.} [\texttt{cond-mat/0509323} and Phys. Rev. A
\textbf{73}, 021603(R) (2006)].  In this Comment we show that both
approaches lead exactly to the same result when the correct
weight function is used in the projection. Plots of the correct
behavior of the phonon and monopole-mode spectra in the BCS,
unitary, and BEC limits are also shown.
\end{abstract}

\maketitle

In their recent study on sound propagation in an elongated Fermi
superfluid in the BEC-BCS crossover, Ghosh and Machida
\cite{Ghosh2006} have reported a calculation of the sound
velocity $u_1$ for the case of a power-law equation of state
(EOS) as previously analyzed for bosons by Zaremba
\cite{Zaremba1998} and for fermions by Capuzzi \textit{et al.}
\cite{Capuzzi2006}. Their result is found to be incorrect due to the
use of an improper projection of the sound wave equation.

The eigenvalue equation for small-amplitude density modes $\delta
n_{q}(\mathbf{r}_{\perp})e^{i q z}$ is obtained by linearization of the
hydrodynamic equations around equilibrium and reads
\begin{multline}
M\omega^2\delta n_q= q^2\left(n_0\left.{\partial \mu}/{\partial n}
\right|_{n=n_0}\!\!\!\delta n_q\right) \\
-\nabla_\perp\cdot\left[n_0\nabla_\perp \left(\left.{\partial
\mu}/{\partial n} \right|_{n=n_0}\!\!\!\delta n_q\right)\right],
\label{eq:collmode}
\end{multline}
where $n_0$ is the equilibrium density profile,
$\omega$ the frequency of the perturbation and $q$ its wave
vector along $z$. Equation (\ref{eq:collmode}) reduces to Eq. (8) in Ref.\
\cite{Ghosh2006} for a power-law EOS $\mu(n)=\mathcal{C}n^\gamma$.

To obtain the dispersion relation $\omega(q)$ for any value of $q$
one must resort to the numerical solution of the eigenvalue equation
(\ref{eq:collmode}). A possible method to solve such an equation
consists of expanding the eigenmodes $\delta n_q$ in a complete set
of basis functions (see \textit{e.g.} Zaremba \cite{Zaremba1998}) as
\begin{equation}
\delta n(\mathbf{r}_{\perp}) = \sum_{\alpha}b_{\alpha}\,\delta
n_{\alpha}(\mathbf{r}_{\perp}),
\end{equation}
where $\alpha=(n_r,m)$ labels the basis functions, with $n_r$ the
radial number and $m$ the number for the azimuthal angular momentum.
By inserting this expression into Eq. (\ref{eq:collmode}) and
projecting the result onto an element of the basis, a matrix
representation of the eigenvalue equation is found, which is
suitable for a numerical solution. This procedure allows  
some freedom in the choice of the basis and of the projection, as long
as these satisfy the boundary conditions. However, in order to obtain a standard
eigenvalue equation of the form 
\begin{equation}
\lambda \mathbf{v} = \mathbb{A}\cdot\mathbf{v}
\end{equation}
one must choose a projection in which the basis is
orthogonal \cite{Morse1953}.  For
the basis functions adopted in Ref.\ \cite{Ghosh2006}, cf.
Eqs.~(10) and  (11), the projection must be performed with a weight
function $w(r) \propto (1-\tilde{r}^2)^{-\gamma_0}$
\cite{Abramowitz1965}, where
$\tilde{\mathbf{r}}=\mathbf{r}_{\perp}/R$ and $\gamma_0=1/\gamma-1$,
$R$ being  the radius of the density profile. The orthogonality
condition thus reads
\begin{equation} \int w(\tilde{r})
\,\delta n_{\alpha}^{*}(\tilde{\mathbf{r}})\,\delta
n_{\alpha'}(\tilde{\mathbf{r}})\,d^2\tilde{r} \propto \delta_{\alpha\alpha'}.
\end{equation}
Therefore, for Eq.\ (13) in Ref.\ \cite{Ghosh2006} to be correct
the integrals defining the matrix $M_{\alpha\alpha'}$ must include the
weight function $w(r)$ and thus read
\begin{align}
M_{\alpha\alpha'} =&  A^2 \int d^2\tilde{r}\,
(1-\tilde{r}^2)^{\gamma_0}\, \tilde{r}^{2+|m|+|m'|}\,e^{i (m-m') \phi}
\nonumber \\
&\times P_{n_r'}^{(\gamma_0,|m'|)}(2\tilde{r}^2-1)\,
P_{n_r}^{(\gamma_0,|m|)}(2\tilde{r}^2-1)
\label{eq:M}
\end{align}
where $P_{n_r}^{(\gamma_0,m)}$ are Jacobi polynomials.
Equation (\ref{eq:M}) is what should have been used in Ref.\
\cite{Ghosh2006}, instead of Eq.\ (14), where the weight function
$w(r)$ is missing and the constant $A$ takes a different value since $\delta
n_{\alpha}(\mathbf{r}_{\perp})$ has not been normalized with $w(r)$.
 A Fermi superfluid with $\gamma=1$ corresponds to a Bose-Einstein
condensate (BEC) of molecules and has been previously analyzed by Zaremba
\cite{Zaremba1998} for bosonic atoms.  In this case $w(r)=1$ and Eq.\
(\ref{eq:M}) reduces to Eq.\ (14) in \cite{Ghosh2006}. 

\begin{figure}
\includegraphics[width=\linewidth,clip=true]{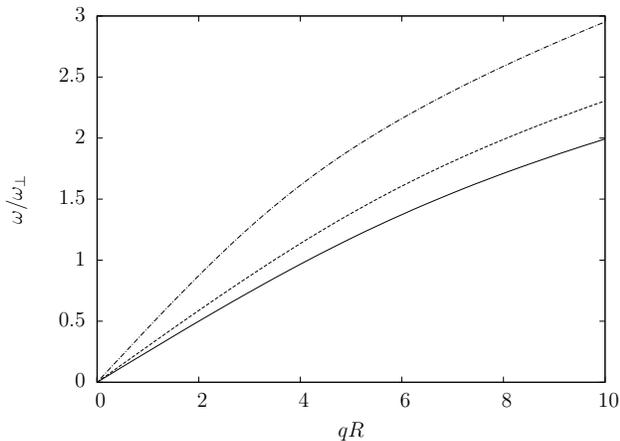}
\caption{\label{fig:cf1}Dispersion relation $\omega(q)$ (in units of
$\omega_{\perp}$) for the lowest-frequency (sound) mode  as a function of $q
R$ with $R$ the radius of the fermion density profile in the BCS
limit. The dash-dotted, dashed, and solid lines correspond to fermions
in the BCS, unitary, and BEC limits, respectively. The BEC limit
corresponds to $y=0.25$ in Eq.\ (26) in \cite{Ghosh2006}.}
\end{figure}

\begin{figure}
\includegraphics[width=\linewidth,clip=true]{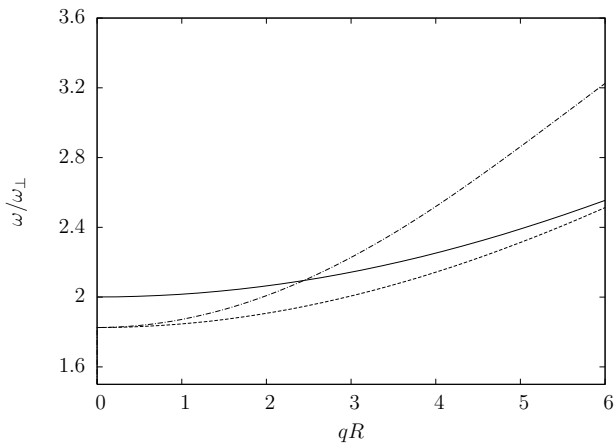}
\caption{\label{fig:cf2}The same as in Fig.\ \ref{fig:cf1} for the monopole mode.}
\end{figure}

To further analyze how the correct orthogonality condition affects the results, we have numerically solved the eigenvalue
equation for sound propagation in a superfluid Fermi gas in the BCS,
unitary, and BEC limits.  Our results for the two lowest frequency 
modes as functions of $q$ are shown in Figs. \ref{fig:cf1} and
\ref{fig:cf2}. The lowest  mode, shown in Fig.\
\ref{fig:cf1}, is sound-like and has a phononic dispersion relation at
long wavelengths. We observe that the slope of the
dispersion relation, \textit{i.e.} the sound velocity, is lower than that found
in Ref.\ \cite{Ghosh2006} and bends down as $q$ increases.
The first excited state, displayed in Fig.\ \ref{fig:cf2}, corresponds 
to a monopolar compressional mode that for $q=0$ is purely radial.
Furthermore its frequency is known
analytically at $q=0$~\cite{Heiselberg2004} as
$\omega_0=\sqrt{10/3}\,\omega_{\perp}$ in the BCS and unitary limits
and as $\omega_0=2\,\omega_{\perp}$ in the BEC limit. Although the
$q=0$ BEC limit of the monopole is correctly quoted in Eq.\ (8) of 
Ref. \cite{Ghosh2006}, it is not correctly depicted in the
corresponding Fig.\ 3, which is to be compared with Fig.\
\ref{fig:cf2} in the present work. We also note that the effective mass associated to this
mode is also different from that found by Ghosh and Machida.

To obtain an analytical expression for the sound velocity $u_1\equiv
d\omega(q)/dq|_{q=0}$ one can use Eq. (13) in Ref. \cite{Ghosh2006}
for the  lowest-frequency mode and expand it to first order in $q^2$.
This demonstrates that the off-diagonal terms in $M_{\alpha\alpha'}$ do
not enter the calculation of the sound velocity, as pointed out by
Zaremba \cite{Zaremba1998} for bosons.  From the definition
(\ref{eq:M}) one obtains $M_{00}=\gamma/(\gamma+1)$ and thus
\begin{equation}
u_1=\sqrt{\frac{\gamma}{2\gamma+2}}v_{\rm F}
\label{chebello}
\end{equation}
with $v_{\rm F}=\sqrt{2\bar{\mu}/M}$ and $\bar{\mu}$ the chemical
potential, in agreement with our result
previously obtained in \cite{Capuzzi2006}.   This is the
sound velocity that also Ghosh and Machida should have obtained if
they had taken into account the weight function $w(r)$.  Their
improper projection of the eigenvalue equation leads them to the
pathological expression $u_1=\sqrt{(2-\gamma)\gamma/4}v_{\rm F}$, which
predicts no sound propagation for $\gamma\ge2$.
The same problem affects the calculation of the effective mass $m_b$ 
for the monopole mode (cf. Eq.\ (27) in Ref.\ \cite{Ghosh2006}), which
once corrected is
\begin{equation}
m_b = \frac{M\,\hbar\omega_{\perp}}{2\mu}\, \frac{(2+2\gamma)^{3/2} 
(1+3\gamma)}{\gamma(1+\gamma+2\gamma^2)}.
\label{eq:mass}
\end{equation}

An alternative and more direct procedure to evaluate the sound
velocity is the one that we have outlined in \cite{Capuzzi2006}. The
spatial dependence is eliminated from Eq.~(\ref{eq:collmode}) above by
integrating in the $(x,y)$ plane. This yields the
dispersion relation 
\begin{equation}
\omega(q) = q\,\left(\frac{1}{M}{\int n_0\,{\partial \mu}/{\partial
  n}|_{n=n_0}\,\delta n_q\,d^2r_{\perp}}\Bigr/{\int \,\delta
  n_q \,d^2r_{\perp}}\right)^{1/2}
\label{eq:dispersion}
\end{equation}
for a perturbation with $\int\delta n_q\,d^2r_{\perp}\neq0$.
Hence, the calculation of the sound velocity requires only the expression of $\delta
n_q$ calculated at $q=0$. By using $\delta n_{q=0} = (\partial
\mu/\partial n|_{n=n_0})^{-1}$ we obtain 
\begin{equation}
u_1 = \left(\frac{1}{M}\int n_0\,d^2r_{\perp}
\Bigr/\int(\partial\mu/\partial n|_{n=n_0})^{-1}d^2r_{\perp}\right)^{1/2}.
\label{eq:u1}
\end{equation}
This expression provides the exact velocity of sound propagation
in cylindrically confined hydrodynamic gases with EOS $\mu(n)$, and
for $\mu(n) \propto n^{\gamma}$ leads to Eq. (\ref{chebello}).

\end{document}